\documentclass{article}
\usepackage{epsf,epic,eepic} 
\usepackage[dvips]{graphicx}
\usepackage{amsmath}
\usepackage{amssymb}

\newcommand {\pa}{\partial}

\newcommand{\rem}[1]{}

\title{\bf{The Zakharov-Kuznetsov Equation as a Two-Dimensional Model for Nonlinear Rossby Waves}}

\author 
{
{\em Georg A. Gottwald} \\
{\em School of Mathematics and Statistics,}\\
{\em  University of Sydney, N.S.W. 2006, Australia}
}
\begin{document}

\maketitle
\begin{abstract}
\noindent We study the dynamics of two-dimensional coherent structures
in planetary atmospheres and oceans. We derive the Zakharov-Kuznetsov
equation for large scale motion from the barotropic quasigeostrophic
equation in a weakly nonlinear, long wave approximation. We consider
coherent structures emerging out of an instability caused by a narrow
jet-like meanflow. We use multiple scale analysis combined with
asymptotic matching.



\end{abstract}

\section{Introduction}
Planetary atmospheres and oceans are strongly turbulent
media. However, highly ordered coherent structures arise in a process
of self-organization, and dominate the dynamics on slow temporal and
large spatial scales.\\
\noindent
The spontaneous appearance of coherent structures is a characteristic
of two-dimensional fluid flows. The basic underlying structure of
these flows is linked to the existence of two quadratic, positive
definite invariants, energy and enstrophy. In spectra of
two-dimensional turbulence one observes two different cascades
associated with these conserved quantities; a direct enstrophy cascade
towards small spatial scales, and an indirect energy cascade towards
larger spatial scales. It is the latter which gives rise to vortex
merging leading to larger and larger vortices. In this Letter we view
the problem as one of weakly nonlinear hydrodynamic stability rather
than turbulence phenomenology.\\
\noindent
Most vortices are monopolar, but dipole and even tripole vortices can
also appear spontaneously. Monopole vortices are mainly created by
shear flow instabilities whereas dipole vortices typically appear when
some additional forcing is applied to the flow.\\
\noindent
The richness and complexity of two-dimensional flows and the
simultaneous presence of motion on very different temporal and spatial
scales makes a direct analysis of the basic equations of motion very
difficult. The presence of rotation reinforces the two-dimensional
character in accordance with the Taylor-Proudman theorem, but rotation
can also introduce baroclinic instability. The latter is a
three-dimensional feature and, thus supports a direct energy cascade
towards small scales. Hence, the dynamics is determined by competing
two-dimensional and three-dimensional processes
\cite{Metais,Bartello,Bartello2,Naulin}. To study vortices in
geophysical fluid dynamics the primitive equations are further reduced
by approximations which allow to focus on temporal and spatial length
scales of vortices \cite{Pedlosky,Zeitlin}. If additionally baroclinic
processes are excluded a further simplification can be made. The
dynamically important variable is the so called potential vorticity
$q$. The resulting quasigeostrophic barotropic vorticity equation
\begin{eqnarray}
\label{charney}
\frac{D}{Dt}q=0 \quad {\rm where} \quad q=\frac{\Delta\psi+f(y)}{H}
\end{eqnarray}
describes large scale motion on a slow time scale. Here $\psi$ is the
stream function, $f(y)$ describes the ambient rotation of the planet
and $H$ is the fluid depth. This equation was first derived by Charney
\cite{Charney}, and then independently by Obukhov \cite{Obukhov}. In
the context of low-frequency drift waves in magnetized plasmas
Equation (\ref{charney}) is known as the Hasegawa-Mima equation
\cite{Hasegawa}. This equation has been the mathematical starting
point for much of the research done on coherent structures and
vortices. It supports so called modons which are localized
soliton-like coherent solutions. Exact modon solutions were obtained
by Larichev \& Reznik \cite{Larichev} for a stationary double-vortex
solution which is antisymmetric in longitude. Extensions to more
general solutions have been made \cite{Flierl,Haines}, and the
spherical geometry of planets has been incorporated
\cite{Tribbia,Verkley,Neven}. However, modons have the drawback that
the potential vorticity is not a smooth function of the stream
function, but may be multivalued.  Therefore interest has grown in
low-dimensional models, although a rigorous proof of existence of a
low-dimensional attractor in quasi\-geo\-strophic systems is still an
unsolved problem. Strictly speaking, one can only define a ``slowest
invariant manifold'' \cite{shepherd}, since the small-scale events,
i.e. the high-frequency and high-wavenumber processes, enlarge the
Hausdorff dimension for the attractor without any convergence
\cite{yano}. Nevertheless, in order to understand better the particular mechanisms
involved in the formation and dynamics of vortices in geophysical
fluid dynamics, it is useful to perform asymptotic techniques to
derive reduced amplitude equations of the basic quasigeostrophic
equations in a multiple scale analysis and study the derived model
evolution-equations. The basic idea is that coherent vortices may be
identified with solitary wave solutions of generic nonlinear
dispersive wave equations.\\
\noindent
Most research has been done in the frame work of the Korteweg-de Vries
equation
\cite{Maxworthy&Redekopp,warn,patoine,hainesandmalanotte,malguzzi1,malguzzi2,mitsudera,JAS1,JAS2}
or in the framework of the Boussinesq equation
\cite{helfrichandpedlosky1,helfrichandpedlosky2}. 
While these models were helpful in describing and identifying
mechanisms for atmospheric blocking, cyclogenesis, meandering of
oceanic streams and the persistence of the Great Red Spot in the
Jovian atmosphere, they are all one-dimensional models with their
obvious limitations.\\
\noindent
In this Letter we will extend weakly nonlinear, long wave multiple
scale analysis to two dimensions and derive the Zakharov-Kuznetsov
equation
\begin{eqnarray*}
A_{T}  + \Delta A_{X} - \mu A A_{X} - \xi A_{XXX} -
\zeta A_{XYY} = 0\;.
\end{eqnarray*}
The Zakharov-Kuznetsov (ZK) equation \cite{ZK} is one of two well-studied
canonical two-dimensional extensions of the Korteweg-de Vries equation
\cite{KdV}; the other being the Kadomtsev-Petviashvilli (KP) equation
\cite{KP}. In contrast to the KP-equation, the ZK-equation has so far
never been derived in a geophysical fluid dynamics context. For a
derivation of the KP equations for internal waves, see
\cite{hydro}. Whereas the KP-equation is valid in isotropic
situations, the ZK-equation is valid in anisotropic settings which is
exactly the case for rotating fluids where the differential
longitudinal dependence of the rotation rate causes anisotropy between
the meridional and the longitudinal directions. Moreover, in contrast
to the KP-equation the ZK-equation supports stable lump solitary
waves. This makes the ZK-equation a very attractive model equation for
the study of vortices in geophsyical flows.\\

\noindent
The Letter is organized as follows. In Section 2 we set up the
barotropic vorticity equation and the mean flow configurations under
consideration. In the beginning of Section 3 we will give a simple
heuristic scaling argument based on the linearized barotropic
vorticity equation to motivate why the ZK-equation is the generic
two-dimensional nonlinear wave equation. In the remainder of Section 3
we will derive the ZK-equation in an asymptotic multiple scale
analysis. Section 4 concludes the Letter with a discussion and an
outlook on further research.

\section{Barotropic Quasigeostrophic Equation}
We shall use a non-dimensional coordinate system, based on a typical
horizontal length scale $L_0$, a typical vertical scale $H_0$, and
typical Coriolis parameter $f_0$. A typical velocity ${\bar U}$ is
taken to be the maximum of the mean current velocity and the time
scale is given by ${\bar U}/L_0$. If we separate the meridional
meanflow $U$ from the perturbation pressure fields $p$ and use the
Boussinesq approximation, we obtain the following equation for the
non-dimensional perturbation pressure field \cite{Pedlosky}
\begin{eqnarray}
\label{qgp}
\left(  \frac{\partial}{\partial t} + U\frac{\partial}{\partial x}
\right)q + \psi_{x}Q_{y} + J \left( \psi,q \right) = 0\; ,
\end{eqnarray}
where
\begin{eqnarray*}
q &=&  \nabla^2 \psi - F\psi \;,\\
\label{Q1}
Q_{y}&=& \beta - U_{yy}+ F U\; ,
\end{eqnarray*}
with Froude number $F$ and the Jacobian defined by $J
\left(a,b\right)= a_xb_y-a_yb_x$. We investigate a channel flow with a
storm track superimposed on a constant meanflow $U_m$ confined at
$y=\pm L$ (see Fig.1). The storm tracks may have a critical layer
where $U(y)=0$. Important is, as we will see, the non-vanishing slope
at at least one boundary of the localized storm track. The boundary
conditions are $\psi = {\rm{const}}$ at $y=\pm \infty$, and we require
that the jet forms a transport barrier to the flow.
\begin{figure}[htb]
  \begin{center}
    \includegraphics[angle=-90,width=.7\textwidth]{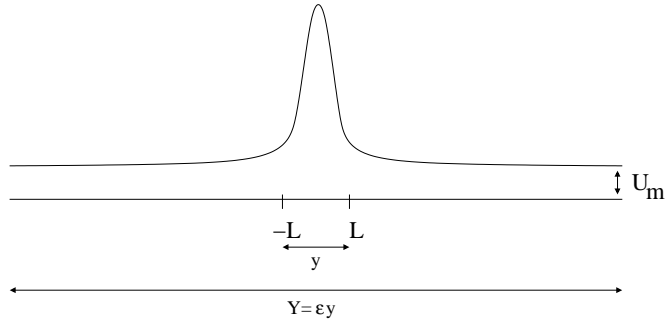}
    \caption{Sketch of a typical mean flow.}
    \label{fig8.1}
  \end{center}
\end{figure}

\vspace{-.4cm}
\section{Nonlinear Wave Equation}
\subsection{Linear Dispersion Relation}
Before we consider the weakly nonlinear, long wave approximation, we
motivate our approach by looking at the linearization of equation
(\ref{qgp}) which yields
\begin{eqnarray}
\label{lin1}
\left( \frac{\partial}{\partial t} + U \frac{\partial}{\partial x}
\right) q + \frac{\partial \psi}{\partial x} \frac{\partial
Q}{\partial y} = 0 \; .
\end{eqnarray}
In terms of $\psi = a_0\exp{i(k(x-ct)+ly)}$ we obtain the dispersion
relation
\begin{eqnarray*}
\label{disp_QG}
c= U-\frac{Q_y}{k^2+l^2+F}\;,
\end{eqnarray*}
provided that the meanflow $U(y)$ is constant. If we focus on
dynamics on the spatial and temporal time scales $X =
\epsilon x, Y=\epsilon y$, i.e. small $k,l$, and $T=\epsilon^3 t$ this is suggestive of
the coupled Zakharov-Kuznetsov equation
\begin{eqnarray}
\label{ZK}
A_{T} + \Delta A_{X} - \mu A A_{X} - \xi A_{XXX} -
\zeta A_{XYY} = 0\;.
\end{eqnarray}
The reason why we expect an equation of the Zakharov-Kuznetsov type
instead of the usual Kadomtsev-Petviashvilli type mostly encountered in
fluid systems is the anisotropic character of (\ref{qgp}) caused by
the $\beta$-effect.
\vspace{-.4cm}
\subsection{Weakly Nonlinear Model}
We consider weakly nonlinear waves riding on a background
meanflow. The meanflow consists of a constant part $U_m$ and a strong
but narrow jetstream (see Fig. 1). The narrow storm track is located
on a short meridional scale $y$. In the outer region the problem
(\ref{qgp}) can be reduced to the linear problem (\ref{lin1}) with
constant meanflow $U_m$. In the interior the structure of the storm
track does not allow for sinusoidal wave solutions but instead we will
derive a nonlinear wave equation. In order for the nonlinear wave
equation which is valid only in the inner region where the storm
tracks are nonuniform, the inner solution has to be matched to the
outer sinusoidal solution.
\subsubsection{Outer solution}
In the outer region where the meanflow is uniform and constant,
(\ref{qgp}) reduces to the simple linear equation (\ref{lin1}) for the
streamfunction with constant coefficients \cite{Pedlosky}.\\ There
the solution of the streamfunction can be written as
\begin{eqnarray}
\label{psi_out}
\psi^{(\rm out)}=a(t,T,X) \sin(lY)+b(t,T,X)\cos(lY)\; ,
\end{eqnarray}
where $l$ is the meridional wave number and is determined by the
dispersion relation of the linearized model (\ref{lin1}).
\subsubsection{Inner solution}
In the interior of the storm track on the small scale $y$, the
meanflow is not constant. We shall study weakly nonlinear long
waves. We introduce the following scales,
\begin{eqnarray*}
\label{tscaling}
&&X = \epsilon x,\hspace{0.7cm} 
Y = \epsilon y,\hspace{0.7cm}
T = \epsilon^3 t,\hspace{0.7cm}\\
&&\psi(X,Y,T,y) = \epsilon^2 \psi^{(0)}  \; + \;
                    \epsilon^3  \psi^{(1)} \; + \;
                    \epsilon^4 \psi^{(2)}  \; + \; 
                    \cdots \; ,
\end{eqnarray*}
where $\epsilon$ is a small parameter, the inverse of which measures the
large scales of the disturbance. Next, we rescale the parameters $F
\rightarrow \epsilon^2 F$ and $\beta \rightarrow \epsilon^2 \beta$.
The scaling of the Froude numbers implies that our model is valid for
situations where the internal Rossby radius of deformation is of the
order of the long horizontal scale. Further, the scaling of $\beta$
implies that $Q_{y}\approx -U_{yy}$ at the lowest order. The boundary
conditions we use are $\psi={\rm constant}$ at $y=\pm \infty$ and
$U_y\psi_{X}=U\psi_{Xy}$ at $y=\pm L$ which simply states that there
is no transport of fluid across the jet-stream.\\ Substituting this
scaling into equation (\ref{qgp}) yields,
\begin{eqnarray*}
0&=&(\epsilon^3\pa_T+\epsilon U\pa_X)
 (\epsilon^4\pa_{XX}\psi^{(0)}
+\epsilon^2\pa_{yy}\psi^{(0)}
+2\epsilon^3\pa_{yY}\psi^{(0)}
+\epsilon^4\pa_{YY}\psi^{(0)}
-\epsilon^4 F\psi^{(0)}\\
&\hphantom{+}&\hphantom{(\epsilon^3\pa_T+\epsilon U\pa_X)}
+\epsilon^3\pa_{yy}\psi^{(1)}
+\epsilon^4\pa_{yy}\psi^{(2)}
+2\epsilon^4\pa_{yY}\psi^{(1)})\\
&+&\epsilon^3\psi_{X}^{(0)}
 (\epsilon^2\beta-U_{yy} + \epsilon^2FU)\; \; 
+ \; \; \epsilon^4\psi_{X}^{(1)}(-U_{yy})\; \; 
+ \; \; \epsilon^5\psi_{X}^{(2)}(-U_{yy})\\
&+&\epsilon^5(\psi_{X}^{(0)}\psi_{yyy}^{(0)}-\psi_{y}^{(0)}\psi_{yyX}^{(0)})
 \; .
\end{eqnarray*}
Counting the orders of $\epsilon$ we obtain to the lowest order,
${\cal{O}}(\epsilon^3)$
%
%
%
\begin{eqnarray}
\label{ord_0}
{\cal{L}} \psi_{X}^{(0)} = 0 \; ,
\end{eqnarray}
where
\begin{eqnarray}
\label{Lop}
{\cal{L}}= \frac{\partial }{\partial y}\left[ U \partial_y - U_{y} \right]\; .
\end{eqnarray}
We look for an amplitude equation, ie we want to write
\begin{eqnarray}
\label{psi0}
\psi^{(0)} (X,Y,T,y) = A(X,Y,T)\varphi^{(0)}(y)
\end{eqnarray}
and seek an evolution equation for the slowly varying amplitude
$A(X,Y,T)$. We easily find
\begin{eqnarray}
\label{varphi0}
\varphi^{(0)}(y) = 
U(y)(1+\int_{-L}^y\frac{\alpha_0}{U^2(y^\prime)}\; dy^\prime) \; ,
\end{eqnarray}
where $\alpha_0$ is a constant of integration. If we allow for zero
meanflow within the narrow jet region we need to impose
$\alpha_0=0$. We summarize the solution of equation (\ref{ord_0})
\begin{eqnarray}
\label{psi_in}
\psi^{(0)} ( X,Y,T,y) = A(X,Y,T) U(y) \; .
\end{eqnarray}
The meridional structure on the small scale $y$ of $\psi$ is
entirely determined by the mean currents at the leading order. \\

\noindent
At the next order, ${\cal{O}}(\epsilon^4)$, we obtain a linear
inhomogeneous equation for $\psi^{(1)}$,
\begin{eqnarray*}
 2U \psi_{yYX}^{(0)} 
+ U \psi_{yyX}^{(1)} 
- U_{yy} \psi_{X}^{(1)} = 0 \; ,
\end{eqnarray*}
%
%
%
which can be written, using (\ref{varphi0}), as
\begin{eqnarray*}
  {\cal{L}} \psi_{X}^{(1)} = -2U\varphi_{y}^{(0)}A_{XY}\; .
\end{eqnarray*}
This equation is again solved by the method of variation of parameters
and we obtain
\begin{eqnarray*}
\psi_{}^{(1)} = \varphi^{(1)} A_{Y}  \; ,
\end{eqnarray*}
with 
\begin{eqnarray*}
\varphi_{}^{(1)}=U(y) 
          (1 - y - L) 
             + \int_{-L}^y\frac{\alpha_1}{U^{2}(y^\prime)}dy^\prime 
\; .
\end{eqnarray*}
We note that the higher order term $\psi^{(1)}$ is slaved to the
$\psi^{(0)}$ term and the dynamics of the corresponding amplitude
equation which will be derived shortly. For the same reasons as above
we set $\alpha_1=0$ and obtain
\begin{eqnarray*}
\psi_{}^{(1)} = U (1 - y - L) A_{Y}  \; .
\end{eqnarray*}
The ${\cal{O}}(\epsilon^5)$ terms give us an evolution equation for
the amplitude $A$. We obtain
\begin{eqnarray*}
&\hphantom{+}&
U \psi_{Xyy}^{(2)} - U_{yy} \psi_{X}^{(2)} 
+ 2U \psi_{yYX}^{(1)}\\
&+&\{ \pa_T\psi_{yy}^{(0)} 
   +U \psi_{XXX}^{(0)}
   +U \psi_{XYY}^{(0)}
   -F U \psi_X^{(0)}\\ 
&+&\psi_{X}^{(0)}\psi_{yyy}^{(0)} 
   -\psi_{y}^{(0)}\psi_{Xyy}^{(0)} 
   +\beta \psi_{X}^{(0)}
   +F U\psi_{X}^{(0)} \}
=0
\; ,
\end{eqnarray*}
which, using (\ref{psi0}), can be written as
\begin{eqnarray}
\label{116}
{\cal{L}}\psi_{X}^{(2)}=-G
\; ,
\end{eqnarray}
where 
\begin{eqnarray}
\label{Gn}
G& =& \varphi^{(0)}_{yy} A_{T}
      + U\varphi^{(0)}(A_{XXX}+A_{XYY})
      -F U
      \varphi^{(0)} A_{X}
\nonumber\\
&+& \varphi^{(0)}
\left(  \beta +FU  \right) A_{X}
+ \left( \varphi^{(0)}\varphi^{(0)}_{yyy}
                  -\varphi^{(0)}_{y}\varphi^{(0)}_{yy}
\right) AA_{X}  
\nonumber\\
&+&2U\varphi^{(1)}_{y}A_{XYY}\; ,
\end{eqnarray}
%
%
%
To assure boundedness of the solutions of (\ref{116}) we have to
require a solvability condition in form of a Fredholm alternative.\\
The homogeneous adjoint problem to equation (\ref{116}) may be written
as
\begin{eqnarray}
\label{adjoint}
{\cal{L}}^{\dagger} \phi =0\; ,
\end{eqnarray}
with 
\begin{eqnarray*}
{\cal{L}}^{\dagger} = 2U_{y}\partial_y+U\partial_{yy}\; ,
\end{eqnarray*}
where we have used the boundary conditions $U_y\psi_X=U\psi_{Xy}$ at
$y=\pm L$. The adjoint eigenvalue problem (\ref{adjoint}) has one
trivial constant kernel mode $\phi_1={\it{const}}$ and one nontrivial,
namely
\begin{eqnarray}
\label{nontriv}
\phi_2(y) =\int_{0}^y\frac{1}{U^2(y^\prime)} d y^\prime \; .
\end{eqnarray}
The nontrivial kernel mode $\phi_2$ has to be discarded because it
does not satisfy the boundary condition. To see this note that the
Fredholm alternative for the elliptic operator \ref{Lop} together with
the boundary condition $U_y\psi_X=U\psi_{Xy}$ is equivalent to one for
the operator
${\cal{L}}^\prime=\partial_y[U\partial_y\psi]-U_y\partial_y\psi-U_{yy}\partial_y\psi$
with the boundary condition $\partial_y\psi=0$ at $y=\pm L$. Also note
that a non-zero kernel mode $\phi_1$ is only consistent with the
boundary conditions if $U(\pm L)\neq 0$.

The solvability condition is thus given by the trivial
constant kernel mode
\begin{eqnarray}
\int_{-L}^L G \; dy = 0 \; .
\end{eqnarray}
On substituting the expressions (\ref{psi0}) with
$\varphi^{(0)}(y)=U(y)$ and (\ref{Gn}) we obtain the desired
amplitude equation for $A$,
\begin{eqnarray}
\label{ZK11}
A_{T} + \Delta A_{X} - \mu A A_{X} - 
 \xi A_{XXX} - \zeta A_{XYY} = 0\; ,
\end{eqnarray}
where
\begin{eqnarray}
\label{parameters}
I &=& -\left[ U_{y} \right]_{-L}^L \; ,\nonumber \\
I\xi &=& \int_{-L}^L U^2 \; dy \; ,\nonumber\\
I\zeta &=& \left[ U^2(1-y-L) \right]_{-L}^L\; ,
\nonumber\\
I \mu &=& - \left[ U_{y}^2 \right]_{-L}^L 
              + \left[ UU_{yy} \right]_{-L}^L \; ,\nonumber\\
I\Delta &=& -\int_{-L}^L \beta U \; dy \; .
\end{eqnarray}
We note that due to the last term of (\ref{Gn}) the Zakharov-Kuznetsov
equation is inhomogeneous in the sense $\xi \neq \zeta$. It is
pertinent to mention that a nonzero $\zeta$ requires a nonzero mean
flow at at least one of the boundaries of the storm track.  The
coefficients of the nonlinear terms $\mu$ require a non-vanishing
slope at at least one boundary. The slope $U_y$ and also $U_{yy}$ at
the boundaries of the jet $y=\pm L$ may be determined from a given
meanflow configuration by averaging over a very short region, say
$y/\epsilon$, where a sudden change of the constant mean flow $U_m$ to
the jet occurs.

\subsection{Asymptotic Matching}
At the lowest order the inner solution (\ref{varphi0}) with
$\alpha_0=0$ and the outer solution (\ref{psi_out}) have to be
matched. The outer solution has been derived on the large scale $Y$
whereas the inner solution and its associated amplitude equation, the
Zakharov-Kuznetsov equation (\ref{ZK11}), were derived on the short
scale $y$. Henceforth we need to require that the asymptotic limit of
the outer solutions for $Y\to 0$ coincides with the asymptotic limit
of the inner solution for $y\to \infty$. The limit of the inner
solution is $\psi^{(in)}=A(X,Y,T)U_m$. The limit of the outer solution
is $\psi^{(out)}=b(X,Y,T)$. Hence we find $b(X,Y,T)=A(X,Y,T)U_m$,
which extends the dynamics of the Zakharov-Kuznetsov equations to the
outer region.\\

\rem{
\noindent
We can perform asymptotic patching at the interface $y=\pm L$ of the
inner and outer solution. Now at the lowest order, the inner solution
(\ref{psi_in}) with $\alpha_0=0$
\begin{eqnarray*}
\psi^{{\rm in}} ( X,Y,T,y) = 
U(y) \; A(X,Y,T) \; ,
\end{eqnarray*}
has to be matched at the boundaries $y=\pm L$ with the outer solution
$\psi_{\rm out}$ (\ref{psi_out}). We treat here for simplicity only one
of the boundaries $y=L$. An analogous calculation has to be performed
for the other boundary at $y=-L$. Moreover we center the solution at
$y=L$. We obtain
\begin{eqnarray*}
\psi^{(\rm out)}=a(t,T,X,Y) \sin(l(Y-L))+b(t,T,X,Y)\cos(l(Y-L))\; .
\end{eqnarray*}
We match $\psi$ and its derivative
\begin{eqnarray}
\label{bcasy}
\psi^{(\rm out)}(L)&=&A(X,Y,T)U(L)\nonumber\\
\frac{d}{dy}\psi^{(\rm out)}(L)&=&A(X,Y,T)\frac{d}{dy}U(L)\;
,
\end{eqnarray}
and obtain 
\begin{eqnarray}
\label{bcsincos}
b=AU(L) \quad {\rm and} \quad a=AU^\prime(L)/l\; .
\end{eqnarray}
Hence the inner solutions, i.e. the solutions of the Zakharov
Kuznetsov equation, also determine the large scale structure of the
outer solution (\ref{psi_out}). Note that we have only required
matching on the small scale $y$ and not on the large scale $Y$ where
the nonuniformity of the inner region is not felt.\\ The case $l=0$ is
possible but requires $U^\prime(L)=0$ which implies that the amplitude
equation (\ref{ZK11}) is purely linear since $\mu=0$ (see
(\ref{parameters})).
}

\section{Discussion}
We have derived the nonlinear dispersive Zakharov-Kuznetsov equation
from the quasigeostrophic barotropic vorticity equation. It is well
known that the ZK-equation, although it is not integrable by means of
the inverse scattering transform, supports a family of steady-shape
stable lump solitary waves, moving at an arbitrary velocity
\cite{ZK,Iwasaki}. These may help to describe two-dimensional
coherent structures such as atmospheric blocking events, long lived
eddies in the ocean or coherent structures in the Jovian atmosphere
such as the Great Red Spot. The model is from an analytical point of
view easier to treat than the full barotropic quasigeostrophic
equation and its solutions do not exhibit multivalued potential
vorticity-stream function relationships as modons do.\\
\noindent
Geophysical flow on large scales is widely accepted to be
conservative. This allows for Hamiltonian descriptions of the flow on
large scales. Our model also exhibits a Hamiltonian structure. Note
that the momentum
\begin{eqnarray*}
P=\int_{-\infty}^{\infty} A^2 \; dX\, dY
\end{eqnarray*}
and the Hamiltonian with the Hamiltonian density
\begin{eqnarray*}
{\cal{H}}=
\frac{\xi}{2}A_{X}^2
+ \frac{\zeta}{2}A_{Y}^2 
- \mu A^3
\; .
\end{eqnarray*}
are conserved.\\
\noindent
We have assumed a meridional meanflow $U$ which consists of a constant
part $U_m$ and a narrow localized storm track. Note that the jet
stream may also be a narrow interface between two regions of meanflow
with opposite flow direction. Such persistent shear layers exist
between the zones and belts in the Jovian atmosphere.\\

\noindent
Analysis of the solutions of (\ref{ZK11}) is planned. Their stability
has to be numerically tested within the Zakharov-Kuznetsov system. The
ZK-equation has been derived using asymptotic techniques and is as
such an asymptotic limit to the barotropic quasigeostrophic vorticity
equation. However, it is not clear that the same is true for the
solutions. The solutions of the Zakharov-Kuznetsov equations do not
necessarily have to be asymptotically close to the solutions of the
full quasigeostrophic system. this is due to the lack of a centre
manifold as discussed in the introduction. In further work we will
test the approximation of the solution numerically by taking solutions
of the ZK-equation and testing their dynamics in the full
quasigeostrophic system.\\

\noindent {\underbar{\bf Acknowledgements }} I would like
to thank Tom Bridges, Daniel Daners, Roger Grimshaw, Charlie
Macaskill, Marcel Oliver, Dmitry Pelinovsky, Victor Shrira and
Vladimir Zeitlin for valuable discussions.


\begin{thebibliography}{99}\label{refs}

\bibitem{Bartello}
P.~Bartello (1995), `{\em{Geostrophic adjustment and inverse cascades in
rotating stratified turbulence}}', J.Atmos.Sci. {\bf
52}, 4410--4428

\bibitem{Bartello2}
P.~Bartello, O.~Metais and J.~Lesieur (1996), `{\em{Geostrophic versus
wave eddy viscosities in atmospheric models}}', J.Atmos.Sci. {\bf
53}, 564--571

\bibitem{shepherd}
O.~Bokhove and T.~Shepherd (1996), `{\em{On Hamiltonian balanced
dynamics and the slowest invariant manifold}}', J.Atmos.Sci. {\bf
53}, 276--297

\bibitem{Charney} J.~Charney (1948), `{\em{On the scale of atmospheric
motions}}', Geophys. Pub. {\bf 17}, 1--17

\bibitem{Flierl} G.~R.~Flierl, V.~D.~Larichev J.~C.~McWilliams and
G.~M.~Reznik (1980), `{\em{The dynamics of baroclinic and baroclinic
solitary eddies.}}, Dyn. Atmos. Oceans {\bf
5}, 1--41

\bibitem{JAS1} G.~A.~Gottwald and R.~H.~J.~Grimshaw (1998), `{\em{The
  formation of coherent structures in the context of
  blocking}}', J.~Atmos.~Sci. {\bf 56}, 3640--3662

\bibitem{JAS2} G.~A.~Gottwald and R.~H.~J.~Grimshaw (1998), `{\em{The
  effect of topography on the dynamics of interacting solitary waves
  as an example for atmospheric blocking}}', J.~Atmos.~Sci. {\bf 56}, 3663--3678
 
\bibitem{hydro} R.~H.~J.~Grimshaw and  Y.~Zhu (1994), `{\em{Oblique
interaction between internal solitary waves}}', Stud. Appl. Math. {\bf 92}, 249

\bibitem{Haines} K.~Haines and J.~Marshall (1987), `{\em{Eddy-forced
coherent structures as a prototype of atmospheric blocking}}',
Quart.J.Roy.Meteor.Soc. {\bf 113}, 681--704

\bibitem{hainesandmalanotte}
K.~Haines and P.~Malanotte-Rizzoli (1991), `{\em{Isolated anomalies in
westerly jet-streams: A unified approach}}', J.~Atmos.~Sci. {\bf 48},
510--526

\bibitem{Hasegawa} A.~Hasegawa and K.~Mima (1978), `{\em{Pseudo-three
dimensional turbulence in magnetized nonuniform plasmas}}',
Geophys. Pub. {\bf 17}, 1--17

\bibitem{helfrichandpedlosky1}
K.~R.~Helfrich and J.~Pedlosky (1993) `{\em{Time-dependent isolated
anomalies in zonal flows}}', J. Fluid Mech.  {\bf 251}, 377--409

\bibitem{helfrichandpedlosky2}
K.~R.~Helfrich and J.~Pedlosky (1995), `{\em{Large-amplitude coherent
anomalies in baroclinic zonal flows}}', J.Atmos.Sci. {\bf 52}, 1615--1629

\bibitem{Ingersoll}
A.~P.~Ingersoll (1990), `{\em{Atmospheric dynamics of the outer planets}}', Science {\bf 248}, 308--315

\bibitem{Iwasaki}
H.~Iwasaki, S.~Toh and T.~Kawahara (1990), `{\em{Cylindrical
quasi-solitons of the Zakharov-Kuznetsov equation}}', Physica D {\bf 43}, 293--303

\bibitem{KP} B.~B.~Kadomtsev and V.~I.~Petviashvilli (1970), `{\em{On
the stability of solitary waves in weakly dispersing media}}',
Sov. Phys. Dokl. {\bf 15}, 539--541

\bibitem{KdV}
D.~J.~Korteweg and G.~de Vries (1895), '{\em{On the change of form of
long waves advancing in a rectangular channel, and a new type of long
stationary waves}}, Phil. Mag. (5) \textbf{39}, 422--443

\bibitem{Larichev} V.~D.~Larichev and G.~M.~Reznik (1976), `{\em{On
two-dimensional solitary Rossby waves}}, Dokl. Akad. Nauk. SSSR {\bf
231},  1077--1079

\bibitem{malguzzi1}
P.~Malguzzi and P.~Malanotte-Rizzoli (1984), `{\em{Nonlinear
stationary Rossby waves on nonuniform zonal winds and atmospheric
blocking. Part I: The analytical theory}}', J.Atmos.Sci. {\bf 41},
2620--2628

\bibitem{malguzzi2}
P.~Malguzzi and P.~Malanotte-Rizzoli (1985), `{\em{Coherent
structures in a baroclinic atmosphere. Part II: A truncated model
approach}}', J.Atmos.Sci. {\bf 42}, 2463--2477

\bibitem{Maxworthy&Redekopp} 
T.~Maxworthy and L.~G.~Redekopp (1976), `{\em{A solitary wave theory
of the Great Red Spot and other observed features in the Jovian
atmosphere}}', Icarus {\bf 29}, 261--271

\bibitem{Metais} O.~Metais, P.~Bartello, E.~Garnier and J.~Lesieur
(1996), `{\em{Inverse cascade in stratified rotating turbulence}}', Dyn. Atmos. Oceans {\bf 23},  193--203

\bibitem{mitsudera}
H.~Mitsudera (1994), `{\em{Eady solitary waves: A theory of type B
cyclogenesis}}', J.Atmos.Sci. {\bf 57}, 734--745

\bibitem{Naulin}
V.~Naulin, K.~H.~Spatschek, S.~Musher and L.~I.~Piterberg (1995),
`{\em{Properties of a two-nonlinearity mode}}', Phys.  Plasmas 2 {\bf
48},  2640--2652

\bibitem{Neven} E.~C.~Neven (1992), `{\em{Quadrupole modons on a
sphere}}', Geophys. Astrophys. Fluid Dyn. {\bf 65}, 105--126

\bibitem{Obukhov} A.~M.~Obukhov (1949), `{\em{On the question on a
geostrophic wind}}', Izv. AN SSSR Geograf. Geofiz. {\bf 13}, 281--306

\bibitem{patoine}
A.~Patoine and T.~Warn (1982), `{\em{The interaction of long,
quasistationary baroclinic waves with topography}}', J.Atmos.Sci.
{\bf 39}, 1018--1025

\bibitem{Pedlosky} J.~P.~Pedlosky (1987),
{\em{ Geophysical Fluid Dynamics}},
Springer-Verlag: New-York


\bibitem{Zeitlin} A.~ Stegner and V.~ Zeitlin (1995), `{\em{What can
asymptotic expansions tell us about large-scale quasi-geostrophic
anticyclonic vortices}}', Nonlin. Proc. in Geophys.  {\bf 2}, 186--193

\bibitem{Tribbia} J.~J.~Tribbia (1984), `{\em{Modons on in spherical
geometry}}', Geophys. Astrophys. Fluid Dyn. {\bf 30}, 131--168

\bibitem{Verkley} W.~T.~M.~Verkley (1984), `{\em{The construction of
barotropic modons on a sphere}}', J. Atmos. Sci. {\bf 41}, 2492--2504

\bibitem{Shrira} V.~V.~Voronovich, V.~I.~Shrira and Y.~A.~Stepanyants
(1998), `{\em{Two-dimensional models for nonlinear vorticity waves in
shear flows}}', Stud. Appl. Maths.  {\bf 100}, 1--32

\bibitem{warn}
T.~Warn and B.~Brasnett (1983), `{\em{The amplification and capture of
atmospheric solitons by topography: A theory of onset of regional
blocking}}', J.Atmos.Sci. {\bf 40}, 28--40

\bibitem{yano}
J.~Yano and H.~Mukougawa (1992), `{\em{The attractor dimension of a
quasi-geostrophic two layer system}}', Geophys. Astrophys.  Fluid
Dyn. {\bf 65},~77--91

\bibitem{ZK} V.~E.~Zakharov and E.~A.~Kuznetsov (1974), `{\em{On
three-dimensional solitons}}', Sov. Phys. JETP {\bf 39}, 285--286

\end{thebibliography}
\end{document}